\documentstyle[12pt]{article}
\newcommand{\beq}{\begin{equation}}
\newcommand{\eeq}{\end{equation}}
\newcommand{\bea}{\begin{eqnarray}}
\newcommand{\eea}{\end{eqnarray}}

\newcommand{\tx}{\textstyle}

\newcommand{\tr}{{\rm tr}}

\newcommand{\vev}[1]{\Big\langle #1 \Big\rangle}

\newcommand{\V}{{\cal V}}
\newcommand{\cO}{{\cal O}}
\input epsf
\begin{document}  

\hfill \vbox{\hbox{UCLA/01/TEP/15}} 
\begin{center}{\Large\bf Vortex waistlines and long range 
fluctuations 
}\\[2cm] 
{\bf Tam\'as G. Kov\'acs}\footnote{Supported by the European 
Community's Human Potential Programme under contract 
HPRN-CT-2000-00145, ``Hadrons/Lattice QCD'' and 
by Hungarian science grant OTKA-T032501.} \footnote{On leave
from Department of Theoretical Physics, University of P\'ecs,
Hungary.} \\
{\em  NIC/DESY Zeuthen, Platanenallee 6,
D-15738, Zeuthen, Germany}\\
{\sf e-mail: kovacs@ifh.de}\\[5mm] 

and\\[5mm]

{\bf E. T. Tomboulis}\footnote{Research supported by 
NSF grant NSF-PHY 9819686}\\
{\em Department of Physics, UCLA, Los Angeles, 
CA 90095-1547}\\
{\sf e-mail: tombouli@physics.ucla.edu}
\end{center}
\vspace{1cm}

\begin{center}{\Large\bf Abstract} 
\end{center} 

We examine the manner in which a linear potential results 
from fluctuations due to vortices linked with the Wilson loop. 
Our discussion is based on exact relations and 
inequalities between the Wilson loop and the vortex and electric flux 
order parameters. We show that, contrary to the customary 
naive picture,  only vortex fluctuations 
of thickness of the order of the spatial linear size of the loop 
are capable of producing a strictly linear potential. 
An effective theory of these long range fluctuations emerges naturally  
in the form of a strongly coupled Z(N) lattice gauge theory. 
We also point out that dynamical fermions introduced in this medium undergo 
chiral symmetry breaking.

\vfill

\pagebreak 

\section{Introduction}

Center vortices are widely believed to be the most important 
degrees of freedom for confinement in Yang Mills theories. (For
a recent review see \cite{CV}.) 
Attempts to isolate and compute the vortex content in the 
path integral at large $\beta$ have been  a very active area of 
study in the last two years. (For a general view of the various 
approaches and issues involved, see \cite{LAT}).

A popular plausibility argument for confinement by vortices goes
as follows. To estimate the vortex contribution to the expectation
of a given Wilson loop, one assumes that typically many thin vortices
--- i.e. thin compared to the physically large $(\gg 1$fm$^2)$ Wilson 
loop area $A$ --- link with 
the Wilson loop. A simple estimate is obtained by subdividing
the surface spanned by the loop into $n=A/a$ pieces of equal area $a$
and assuming that a vortex piercing a given small area
piece is present with probability $p$. Provided that the vortices
linking with the loop at different places are {\em independent},
the Wilson loop expectation is then given by
\beq
 \sum_{k=0}^n \left( \begin{array}{c} n \\ k \end{array} \right)
    (-1)^k p^k (1-p)^{n-k} = (1-2p)^{A/a},
       \label{eq:vor_exp}
\eeq
which is the desired area-law.

In the present paper we point out that while this simple picture 
indeed gives the correct area law, it is fundamentally incorrect 
because it is not supported by the underlying full non-Abelian gauge
theory. The response of the vortex distribution 
in the vacuum to the introduction of the external probe represented 
by a Wilson loop should be such as to minimize the effective 
quantum action (free energy) of the system. To examine this we make 
the above arguments more precise
by introducing a set of vortex containers linking with the Wilson
loop in the spirit of \cite{MP}. Any set of vortex containers
can then be used to obtain an upper bound to the Wilson loop by 
making use of rigorous inequalities mostly based on reflection positivity.  
We show that thick vortex containers of thickness of the order of 
the (shorter) side length of the Wilson loop yield the
`best' and in fact the only strict area-law upper bound.
This shows that vortices do not have some fixed characteristic thickness but
the most important vortex fluctuations disordering a Wilson loop
of a given size have a core thickness of the order of the 
linear size of the Wilson loop. 
It also means that generally vortices  cannot  be unambiguously 
located as  individual objects in every given single  
gauge field configuration. The only meaningful quantity is the number 
of vortices (mod $N$) linking with a given Wilson loop, and 
this is a well-defined gauge invariant quantity. 
In the remainder of the paper we show how the physical picture implied 
by the previous discussion leads     
to a simple effective $Z(N)$ 
gauge theory for the long distance center degrees of freedom.
We then observe that the introduction of dynamical fermions in 
this effective strongly coupled system results in chiral symmetry 
breaking. 

The paper is organized as follows. In Section 2 we review and extend 
various relations and inequalities relating the electric free energy 
(Fourier transform of the vortex free energy) to the Wilson loop. 
These relations form the basis for our discussion of vortices and 
the emergence of a linear potential in Section 3. Section 4 
further discusses the physical picture and 
introduces the effective theory. Section 5 contains some conclusions.

\section{Electric flux inequalities}

After a brief description of our notation, in this Section
we introduce the basic electric flux inequalities. These are
already interesting in their own right since they yield a rigorous
upper bound on the Wilson loop in terms of the electric flux 
free energy. 

We work on a hypercubic lattice $\Lambda$ of length $L_\mu$ in 
spacetime direction $\mu= 1,\ldots, d$. We assume the standard
Wilson formulation of lattice gauge theory with $SU(2)$ group-valued
link variables, the Wilson action and periodic boundary conditions
in all directions. Expectation values of observables are defined as
\beq
 \langle \cO \rangle = \frac{1}{Z} \int d[U] \, \cO \, 
        \exp \Bigg(\,\beta \sum_p \frac{1}{2}\mbox{tr} U_p\,\Bigg),
\eeq 
where the integration is over all the group-valued link variables
and $Z$, the partition function, is the same integral without the
operator insertion $\cO$.

Let us denote by $\cO[\V_{\mu\nu}]$ the operator that flips
the sign of the coupling (introduces a $Z(2)$ `twist') 
on a coclosed $(d-2)$-dimensional set of
plaquettes $\V_{\mu\nu}$ winding around the periodic lattice
in the directions perpendicular to the $\mu\nu$-directions.  
The expectation value of this operator
defines the vortex free energy: 
\beq 
\exp (-F_v[\mu\nu]) = \vev{\cO[\V_{\mu\nu}]} \label{vfe}.
\eeq 
The twist amounts to a discontinuous gauge transformation with 
multivaluedness in $Z(2)$, i.e. forces the presence of 
a $\pi_1(SU(2)/Z(2))$ vortex 
wrapped around the periodic lattice. 
As indicated by the notation, the expectation depends 
only on the directions in 
which $\V$ winds through the lattice, not the exact shape or location of 
$\V$. This expresses the mod 2 conservation of flux. 
Indeed, the twist $-1$ on the plaquettes forming 
$\V$ can be moved to the plaquettes forming any other homologous 
coclosed set $\V^{\ \prime}$ by the change of variables 
$U_b \to -U_b$ in the numerator in (\ref{vfe}) for each bond $b$ 
in a set of bonds cobounded by $\V \cup \V^{\ \prime}$. By the same token 
(\ref{vfe}) is invariant under changes mod 2 in the number of 
homologous twisted coclosed sets introduced in $\Lambda$. 
A simple consequence of this is that 
\beq
\langle \cO[\V_{\mu\nu}] \, \cO[\V^{\ \prime}_{\mu\nu}] \rangle = 1.
   \label{mod2}
\eeq

We will assume that, for sufficiently large $|A_{\mu\nu}|$, 
and dimension $d\leq 4$, the vortex free energy (\ref{vfe}) 
behaves as 
\beq
F_v[\mu\nu] \sim  (\prod_{\lambda\not= \mu\nu} L_\lambda\,) 
\;\exp(\,- \rho(\beta)\,|A_{\mu\nu}|\,),  \label{vfe1}
\eeq
where $|A_{\mu\nu}| = L_\mu L_\nu$. This is the optimal  
behavior under exponential transverse spreading (creation 
of mass gap) of the flux introduced by the twist on $\V$, with 
$\rho$ approaching, at least asymptotically, the exact 
linear potential string tension. This behavior is  
expected by physical reasoning \cite{tH}, and explicitly seen in the 
strong coupling expansion \cite{Mu}. Recently, it became possible to   
demonstrate this in numerical simulations at large $\beta$ \cite{KT1}, 
\cite{HRR}. The behavior (\ref{vfe1}) for a `vortex in a box' 
is essential for our argument in the following.

The $Z(2)$ Fourier transform of (\ref{vfe}), 
\beq
\exp(-F_{\rm el}) = \vev{\,\frac{1}{2}(\, 1 - \cO[\V_{\mu\nu}]\,)}, 
\label{efe}
\eeq 
gives the corresponding dual (w.r.t. the center) color-electric     
free energy. The mod 2 conservation of the magnetic flux 
is now expressed by the projection property
\beq
\vev{\,\frac{1}{2}(\, 1 - \cO[\V_{\mu\nu}]\,)
\frac{1}{2}(\, 1 - \cO[\V_{\mu\nu}^{\ \prime}]\,)}
   \; = \; \vev{\,\frac{1}{2}(\, 1 - \cO[\V_{\mu\nu}]\,)}, 
\label{proj}
\eeq 
as is easily seen by using  
Eq.\ (\ref{mod2}).

Consider now a rectangular Wilson loop $C$ placed, say, 
in the $[12]$-plane. Let $\V$ be a coclosed stack of 
plaquettes winding around the periodic lattice in the 
perpendicular $\mu=3,\ldots,d$ directions and through $C$, and  
insert unity in the numerator in the Wilson loop expectation  
$W[C]=\vev{\,{1\over2}\tr U[C]}$ in the form 
\[ 1 = {1\over 2}(\,1 -\cO[\V] \,) 
+ {1\over 2}(\, 1 + \cO[\V]\,)\, . \] 
Then 
\bea 
W[C] &=& \vev{\,\frac{1}{2}\tr U[C]\; \frac{1}{2}
(\,1 - \cO[\V]\,)\,} + 
\vev{\,\frac{1}{2}\tr U[C]\; \frac{1}{2}
(\,1 + \cO[\V]\,)\,} \nonumber \\
  & =& \vev{\,\frac{1}{2}\tr U[C]\; \frac{1}{2}
(\,1 - \cO[\V]\,)\,} + 
\vev{\,\frac{1}{2}\tr U[C]\; \frac{1}{2}
(\,1 - \cO[\V^{\ \prime}]\,)\,} \label{W2}
\eea
Here $\V^{\ \prime}$ is another coclosed stack of plaquettes 
winding around the lattice in the perpendicular 
$\mu=3,\ldots,d$ directions but not threading through $C$. 
The second equality in (\ref{W2}) is obtained by making the  
change of variables $U_b \to -U_b$ in the second term in the 
first equality in (\ref{W2}) for each bond $b$ in a set of bonds 
cobounbed by $\V$ and $\V^{\ \prime}$. This `moves' the twist (-1) on 
the plaquettes forming $\V$ to those forming $\V^{\ \prime}$. 
This set of bonds necessarily involves one (or an odd number of) 
bond(s) on $C$, which  results in the minus sign in the second 
term in the second equality. (\ref{W2}) is represented graphically as 
\beq
\begin{minipage}{9cm}
\epsfysize=1cm\epsfbox{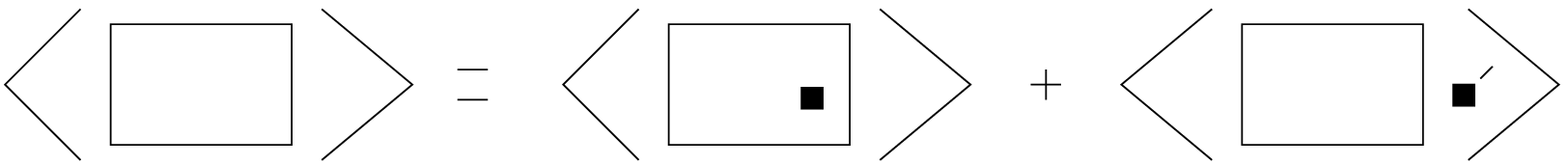}
\end{minipage} 
\label{W2a}
\eeq
where a filled square stands for the operator 
$\frac{1}{2}(\,1-\cO[\V]\,)$, with $\V$ crossing the two-dimensional 
plane containing the loop at  the location of the filled square, and 
winding around the lattice in the remaining $d-2$ perpendicular directions.  

Simple identities like (\ref{W2a}), or (\ref{W4a}), (\ref{W4b}) below, 
serve as the starting point for deriving relations 
between the Wilson loop and vortex free energies 
by use of reflection positivity. 
Given a reflection $r$ about a $(d-1)$-dimensional lattice hyperplane,  
one defines an antilinear mapping $\theta$ on functions of the 
bond variables by $\theta F[\{U_b\}] = F^*[\{U_{rb}\}]$. By the 
reflection positivity properties of the LGT action \cite{RP}, this 
induces a positive semidefinite inner product on the space of 
configurations allowing the use of the corresponding Schwarz 
inequalities. Thus, starting with (\ref{W2a}), consider a reflection 
about a $d-1$-dimensional hyperplane $\pi$
perpendicular to the `vertical' loop legs and containing, 
say, the top `horizontal' leg of the loop C. One then has the 
inequality 
\beq
\begin{minipage}{15cm}
{\epsfysize=2cm\epsfbox{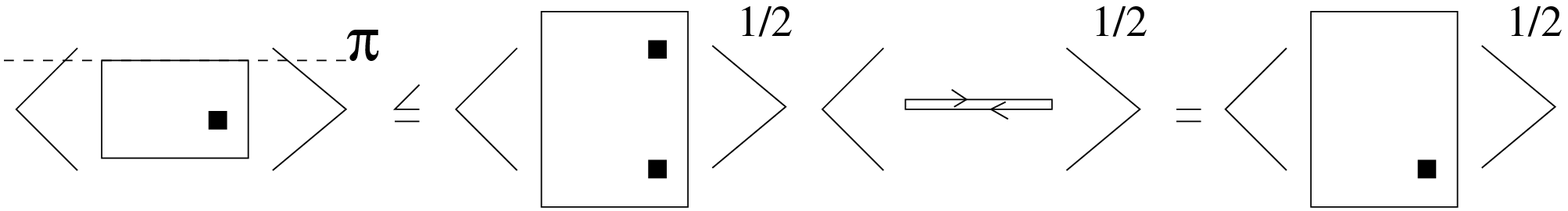}\hfill}
\end{minipage} 
\label{schwarz}
\eeq 
for the first term, and a similar statement for the second term in 
(\ref{W2a}). In (\ref{schwarz}) we made use of (\ref{proj})\footnote{
Notice that (\ref{proj}) remains valid in the presence of additional
operators in the expectation, as long as moving $\V^{\ \prime}$ to
$\V$ by a change of variables does not affect the additional operator.
In particular, this is the case for a Wilson loop when neither or both 
$\V$ and $\V^{\ \prime}$ link with the loop.\label{fproj}} and 
$\frac{1}{2}\tr {\bf 1}=1$. The loop has now doubled in size along the 
direction of one 
of its legs. Proceeding now by repeated reflections in 
hyperplanes containing one of the legs of the loop resulting from the 
previous reflection, and use of (\ref{proj}), one may eventually 
completely eliminate all $\tr U$ factors from the loop operator 
by virtue of the lattice periodicity. Applying this procedure to
both terms on the rhs.\ of Eq.\ (\ref{W2a}) one obtains \cite{TY} 
\beq 
\vev{W[C]} \leq 2 \Bigg(\,\exp(-F_{\rm el})\,\Bigg)^{|A_C|/|A_{12}|}, 
\label{ineqI}
\eeq 
where $|A_C|$ is the minimal area bounded by the Wilson loop.
If then the vortex free energy behaves as in (\ref{vfe1}), 
(\ref{ineqI}) implies area law for the Wilson loop. 

\begin{figure}[hbt]
{\hfill\epsfysize=5cm\epsfbox{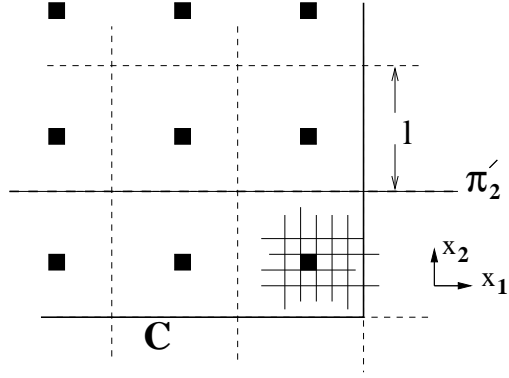}\hfill}
\caption[vg4]{\label{vg4}Arrangement for reflections in 
first term in (\ref{W3}) to obtain (\ref{chess})(see text).}
\end{figure}

Note that the result (\ref{ineqI}) manifestly incorporates mod 2 
conservation since of course (\ref{efe}) and (\ref{vfe}) do. 
This is an important point that we now explore a bit further. 
Any multiple factors of $\frac{1}{2}(\,1-\cO[\V]\,)$ occurring in the 
derivation above were eliminated by (\ref{proj}). 
Suppose instead that we keep a number 
of such factors to make contact with the naive picture of a Wilson loop 
pierced by several independent vortices. 
So imagine that we subdivide the 2-dimensional plane $A$ containing $C$ 
into large squares of side length $l$, i.e. we view $A$ from a coarse 
lattice of spacing $l$. We denote $A$ by $A^\prime$ when viewed from 
the coarse lattice. 
Rewrite the 
identity (\ref{W2a}) using (\ref{proj}) in the equivalent form 
\beq
W[C] = \vev{\,\frac{1}{2}\tr U[C]\; \prod_i\frac{1}{2}
(\,1 - \cO[\V_i]\,)_i\,} + 
\vev{\,\frac{1}{2}\tr U[C]\;\prod_j \frac{1}{2}
(\,1 - \cO[\V^{\ \prime}_j]\,)_j\,} \label{W3} 
\eeq
where the product in the first term includes one factor (indexed by $i$) 
for every large square in $A$ ( every plaquette in $A^\prime$)  
tiling $C$ (figure \ref{vg4}), 
and the identical arrangement 
translated outside $C$ in the second term. 
One now applies reflection positivity to repeatedly reflect about 
$(d-1)$-dimensional hyperplanes $\pi_1^\prime$, or $\pi_2^\prime$, 
perpendicular to the $\mu=1$, or $2$-direction, respectively, 
containing bonds in $A^\prime$ (figure \ref{vg4}). In this way 
one derives 
\beq 
W[C] \leq 2 \prod_i \;\vev{\,\prod_{k\in A^{^\prime}} 
\frac{1}{2}(\,1 - \cO[\V_k]\,)_i\,}^{1/|A^\prime|} 
\label{chess}
\eeq
where in the product inside the expectation there is one factor 
for each plaquette in $A^\prime$.  
(\ref{chess}) now has the form of a typical `chessboard estimate' 
inequality (see e.g. \cite{RP}, \cite{TY}). It is of course equivalent 
to (\ref{ineqI}) by (\ref{proj}) since $|A^\prime|=|A|/l^2$, and 
the number of factors in the outside product equals $|A_C^\prime|= 
|A_C|/l^2$.

Returning  to (\ref{W2}) and noting that 
\bea
& & \vev{\,{\tx\frac{1}{2}}\tr U[C]\; {\tx\frac{1}{2}}
(\,1 - \cO[\V]\,)\,} + 
\vev{\,{\tx\frac{1}{2}}\tr U[C]\; {\tx\frac{1}{2}}
(\,1 - \cO[\V^{\ \prime}]\,)\,} \nonumber \\
&= & \vev{\,{\tx\frac{1}{2}}\tr U[C]\; 
{\tx\frac{1}{2}}(\,1 - \cO[\V]\,)\,{\tx\frac{1}{2}}(\,1 
+ \cO[\V^{\ \prime}])\,}  + 
\vev{\,{\tx\frac{1}{2}}\tr U[C]\; 
{\tx\frac{1}{2}}(\,1 + \cO[\V]\,)\,{\tx\frac{1}{2}}(\,1 
- \cO[\V^{\ \prime}]\,)\,}
\nonumber \\ 
&= & \vev{\, {\tx\frac{1}{2}}\tr U[C]\; 
{\tx\frac{1}{2}}\Big(\,1 - \cO[\V]\;\cO[\V^{\ \prime}]\,\Big)\,}, 
\label{W4}\\
&= & \vev{\, {\tx\frac{1}{2}}\tr U[C]\; 
{\tx\frac{1}{2}}\Big(\,1 - \cO[\V^{\ \prime\prime}]\,\Big)\,}
\label{W4o}
\eea 
one obtains the alternative identities represented graphically by 
\bea
W[C] & = & \begin{minipage}{12cm}
\epsfysize=2.5cm\epsfbox{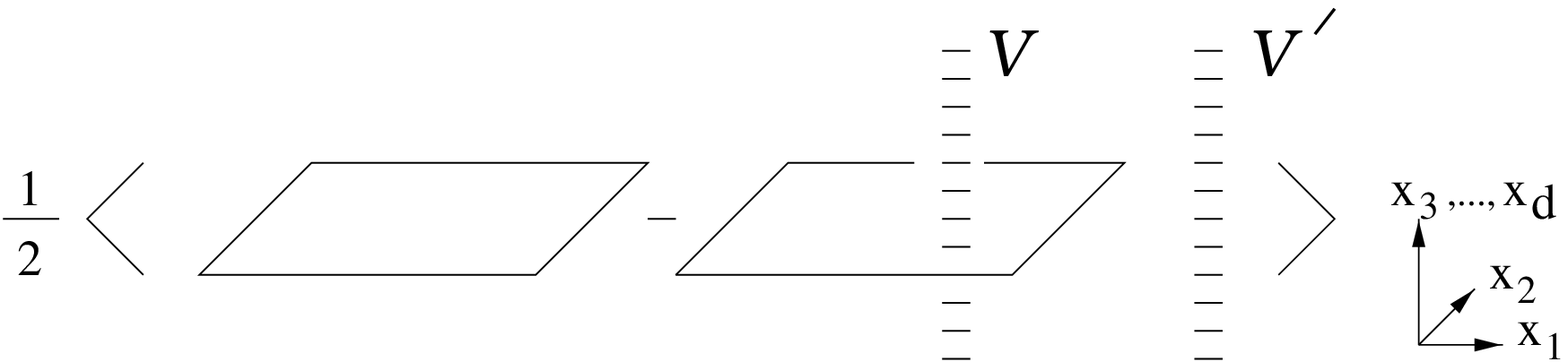}
\vspace{.5cm}
\end{minipage} \label{W4a}\\
   & = & \begin{minipage}{12cm}
\epsfysize=2.2cm\epsfbox{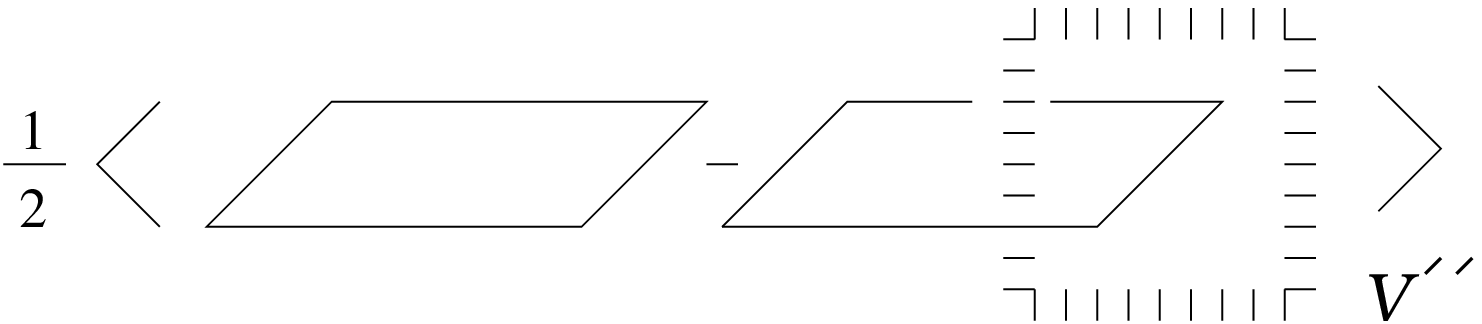}
\end{minipage} \label{W4b}
\eea  
(\ref{W4a}) is the graphical representation of (\ref{W4}). 
(\ref{W4o}), graphically depicted in (\ref{W4b}), is then obtained 
from (\ref{W4}) by simply 
merging $\V$ and $\V^{\ \prime}$ in the  coclosed set 
$\V^{\ \prime\prime}$ linking with $C$ by another shift in integration 
variables and mod 2 flux conservation.\footnote{$\V\cup\V^{\ \prime}
\sim \V^{\ \prime\prime}$ is topologically trivial w.r.t. 
the lattice $T^d$ topology, but not w.r.t. the obstruction 
of the loop $C$.}

Alternatively, (\ref{W4a}), (\ref{W4b}) may be directly obtained 
as follows. Insert  $1=\int_{Z(2)} \,d\gamma$, where $\gamma \in Z(2)$, 
in the numerator in the expectation $W[C]$, and make a shift of 
integration variables $U_b \to \gamma\,U_b$ for all $b\in B$, where 
$B$ is a set of bonds whose coboundary is $\V\cup\V^{\ \prime}$ or 
$\V^{\ \prime\prime}$. The result is (\ref{W4a}), 
or (\ref{W4b}), respectively.

\section{The Wilson loop in terms of vortex containers}

In this Section we introduce a set of vortex containers linking
with a given Wilson loop. Using the relations derived in
the previous Section, we obtain an upper bound on the Wilson 
loop in terms of vortex fluctuations occurring in the given set of
vortex containers. We then ask the question, which set of vortex
containers gives the ``best'' (i.e.\ the lowest) upper bound 
to the Wilson loop. It turns out that the favored set of 
vortex containers has only one single thick container utilizing
an area of the order of the area spanned by the Wilson loop. 
This yields a strictly linear
lower bound to the heavy quark potential, whereas a collection
of many thin vortex containers results in a suppression of the 
potential by a logarithmic factor.

\begin{figure}[hbt]
{\hfill\epsfysize=5cm\epsfbox{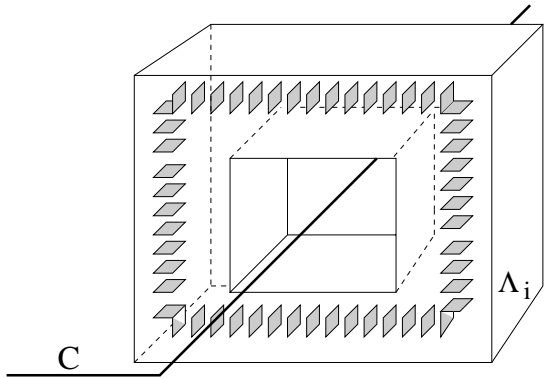}\hfill}
\caption[vg7]{\label{vg7}Vortex container $\Lambda_i$  
enclosing coclosed set of plaquettes $\V_i$ (shaded) linking with 
Wilson loop $C$. $d=3$, or 3-dimensional section in $d=4$.}
\end{figure}

By (\ref{proj}) and footnote \ref{fproj}, one can insert multiple 
$\frac{1}{2}(\,1-\cO[\V]\,)$ factors in (\ref{W4o}), (\ref{W4b}) 
corresponding to a collection of coclosed sets $\{\V_i\}$ linking with 
the loop $C$. Imagine enclosing each $\V_i$ in a `vortex container' 
\cite{MP}, 
i.e. a sublattice $\Lambda_i\subset \Lambda$ containing $\V_i$ and 
wrapping around $C$ (figure \ref{vg7}). 
Imagine integrating over 
the bond variables 
in the interior of each container, keeping bond variables on its 
boundary $\partial\Lambda_i$ fixed, and define  
\beq 
f_{\Lambda_i}(U_{\partial\Lambda_i}) = \frac{1}{2}\,\Bigg(\, 1 - 
{z^{(-)}_{\Lambda_i}(U_{\partial\Lambda_i})\over 
z^{(+)}_{\Lambda_i}(U_{\partial\Lambda_i})} \,\Bigg) ,\label{vcon}
\eeq
where 
\beq
z^{(\mp)}_{\Lambda_i}(U_{\partial\Lambda_i}) \equiv 
 \int\prod_{b\,\in \,\Lambda_i-\partial\Lambda_i} dU_b \;
\prod_{p\,\in \,\Lambda_i-\partial\Lambda_i} \exp (\,\frac{\beta}{2} 
\,(\mp1)^{\V_i[p]}\,\tr U_p\,)\label{pfcon}
\eeq 
with the characteristic function $\V_i[p]\equiv 1$ if $p\in\V_i$, $0$ 
otherwise. $z^{(+)}_{\Lambda_i}(U_{\partial\Lambda_i})$ is 
of course simply the partition function  
for (the interior of) $\Lambda_i$. Then one obtains \cite{MP}:  
\beq 
W[C] = \vev{\;{\tx\frac{1}{2}}\tr U[C]\;
\prod_i\,f_{\Lambda_i}(U_{\partial\Lambda_i})\;}.\label{W5}
\eeq

(\ref{chess}) can be rewritten in the same way: 
\beq
W[C] \leq 2 \prod_i \;\vev{\,\prod_{k\in A^{^\prime}} 
\,f^i_{\Lambda_k}(U_{\partial\Lambda_k})
\,}^{1/|A^\prime|} ,
\label{chess1}
\eeq
where now each container $\Lambda_k$ containing the coclosed 
set $\V_k$ is of transverse area $l^2$ and wraps around the 
lattice in the longitudinal directions. 

Note that $f_{\Lambda_i}(U_{\partial\Lambda_i})$, (\ref{vcon}), 
is nothing but (\ref{efe}) ($Z(2)$ FT of vortex free energy) 
now defined on a lattice $\Lambda_i$ (the vortex container) 
with fixed (instead of periodic) b.c.  
in the transverse directions, but still periodic b.c. in the 
longitudinal directions (in which the $\Lambda_i$'s 
have torus topology by construction). Vortex containers including  
integration over fluctuations (summation of entropy effects) 
in their interior are a convenient device for discussing scales 
larger than their thickness in terms of free energy 
costs rather than the action of individual configurations. 

From (\ref{W5}) one can now, trivially, obtain the bound: 
\beq
W[C] \leq \prod_i\; \max_{U_{\partial\Lambda_i}}
|f_{\Lambda_i}(U_{\partial\Lambda_i})|, \label{ineqII}
\eeq 
where the maximum is taken over all values of the bond variables 
on the boundary $\partial\Lambda_i$.

(\ref{ineqII}) makes a direct connection with the naive 
notion of independent vortices winding 
through a large Wilson loop, resulting in disorder, and 
area law, provided they are able to grow sufficiently long to pierce 
through the loop at any point.  
For this to be possible the vortices must be allowed to grow sufficiently 
thick to keep their free energy cost fixed as their length increases
with that of the Wilson loop. Let $d_i$ be the size of $\Lambda_i$ 
in each of the two directions transverse to the set $\V_i$ used in 
its definition; its longitudinal size is given by $|\V_i|$. 
We must first assume that all the vortex containers are 
thick enough to reach the regime where 
\beq 
 -\ln(\,z^{(-)}_{\Lambda_i}(U_{\partial\Lambda_i})/ 
 z^{(+)}_{\Lambda_i}(U_{\partial\Lambda_i})\,) 
 \; \sim \; |\V_i|\exp(-\rho\,d_i^2)
     \label{eq:free_E}
\eeq
for all $U_{\partial\Lambda_i}$. So to keep the free energy cost
of each vortex less than a fixed value $f$, we need
\beq
d^2_i \geq \frac{1}{\rho} \; \ln\left( \frac{|\V_i|}{f} \right). 
\eeq
Then also $\max f_{\Lambda_i}(U_{\partial\Lambda_i}) < \frac{1}{2}
(1-e^{-f}) < {\rm const}$. But $|\V_i|$ (in $d$ dimensions) 
is of the order $R^{d-2}$ for 
linkage through points away from the perimeter of   
a rectangular loop of side  
lengths $T$ and $R$, $T >  R$. Such a loop 
can then accommodate $\sim RT/\ln R$ 
containers wrapped around it. Thus (\ref{ineqII}) 
gives a confining but not quite purely linear potential 
\beq 
V(R)  \geq \mbox{const}\,R/\ln R \,. \label{quasilin}
\eeq  
   
The same reasoning, and consequent failure to produce a purely 
linear potential, applies  
to the familiar argument for  
confinement by vortices, outlined in the introduction,   
which tacitly underlies or is implied by many discussions in 
the literature.  
One assumes   randomly distributed vortices of 
a certain thickness and basically arbitrary length. 
It is crucial that one assumes that the cost and hence the 
probability for vortices to link anywhere with the loop is fixed 
for any Wilson loop size (cp. (\ref{eq:vor_exp})).
One then considers one vortex linked with a large Wilson loop. 
With the vortex thickness assumed much less than 
the loop's linear dimensions,  
one now sums over all positions of intersection with a 
surface spanning the loop. With the above assumptions, this produces 
a factor proportional to the loop area. 
One then sums over all intersection 
points for two {\it independent} vortices linking with 
the loop, and so on. This clearly exponentiates 
generating area law:
\beq
W[C] \sim 1 + (-1)\, k|A_C| + {((-1)\,k|A_C|)^2\over 2!} + \cdots =
\exp(\,- k|A_C|\,)\;. \label{naive}
\eeq  
We now see that purely linear confinement is obtained this way only 
by adopting a non-interacting gas picture, and ignoring the actual 
free energy requirements for having vortices of 
sufficient length link anywhere with a large loop:  
the type of discussion just given above for (\ref{ineqII}) applies 
to each term in such a summation. Thus, for one vortex linking with 
the large loop of side lengths $T$ and $R$ ($T \gg R$), 
a vortex cross section area of order $\ln R$ is required; 
otherwise, linking anywhere far away from the perimeter 
for fixed, bounded vortex free energy cost $f$, 
as required by the argument, is not possible. 
This leads at best to (\ref{quasilin}), {\it not} (\ref{naive}).

The problem arises because one treats the vortices 
as localized and independent. For sufficiently thick 
vortices free energy costs are indeed 
correctly estimated in magnitude as above, i.e. (\ref{eq:free_E}). 
Thus, if one imagines 
each vortex enclosed in a vortex container of fixed, but 
sufficiently large, width $d$, the exponential transverse 
spreading $\sim \exp(-\rho\,d^2)$ renders the overall 
vortex bulk free energy cost inside insensitive to the exact values 
of the gauge fields on the container boundary. 
The vortex, however, is surrounded by the pure gauge long tail that 
encodes its nontrivial topology, and flux quantization. 
This tail incurs no additional action cost, but is of infinite range  
and communicates the presence of nontrivial topological flux inside the 
container to everywhere outside. So even though the 
gauge field values on a thick container's boundaries are irrelevant for 
estimating the bulk free energy cost inside, they are very much 
relevant for signalling the presence of a vortex inside 
to other vortices or other topological obstructions outside. 

\begin{figure}[hbt]
{\hfill\epsfysize=6cm\epsfbox{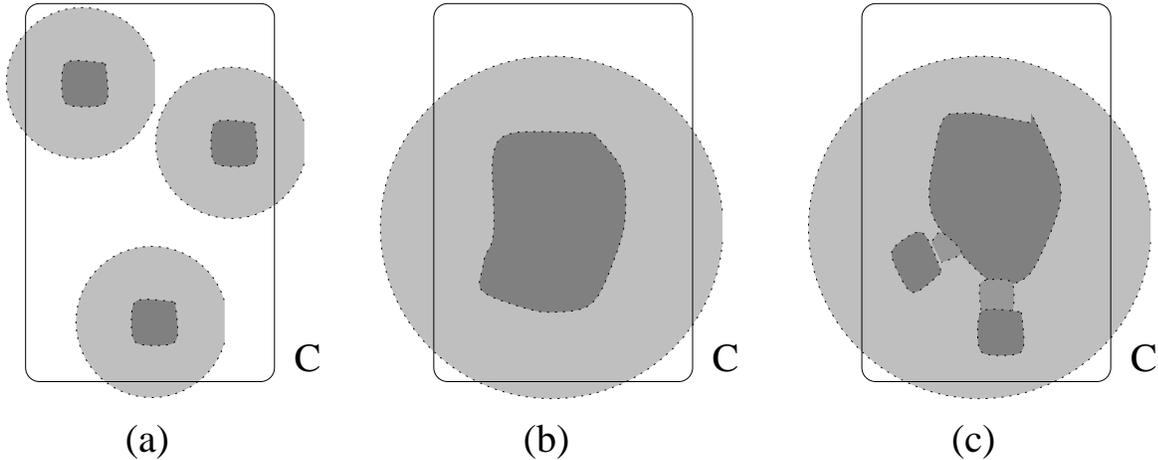}\hfill}
\caption[vg8a]{\label{vg8a}Vortices linking with Wilson loop $C$. 
Cores shown in darker shading, and long range (infinite extent) 
pure gauge tails in light shading:  
(a) Well-separated vortices of approximately fixed width; 
(b) Configurations of equivalent mod 2 flux having one thickened  
vortex lowering potential energy; (c) Configurations contributing 
essentially as in (b) showing the possible ambiguities in counting 
vortices beyond mod $N$.}
\end{figure}

This acts like an `irreducible' interaction between vortices that 
acts at all distances, and enforces flux conservation mod $N$. 
This interaction allows a system of vortex excitations  
to adjust the amount of flux spreading, i.e adjust the 
thickness of vortices to minimize the free energy of the 
system. The thickness of 
vortex cores then is not fixed, but is adjusted relative to 
their length as required by the presence of other vortices and/or  
other obstructions (e.g. Wilson loop legs) sensitive   
to the presence of topological $Z(N)$ flux (figure \ref{vg8a}).  
This means that in general 
vortices cannot be considered isolated, and a definite 
number of vortices, specified more precisely than mod $N$, 
cannot necessarily be unambiguously assigned to every configuration. 

Thus, in the presence of the Wilson loop source, the optimal 
configurations for the system are not those of multiple isolated  
linked vortices, each of some fixed free energy 
cost (figure \ref{vg8a}(a)), hence length 
$|\V| \sim R^{(d-2)}$ and fixed width $d^2 \sim \ln R \ll R,T$. 
It is more advantageous, in terms of free energy cost, for 
multiple linking vortices to thicken and merge,  
the total topological flux being conserved mod N  
(figure {\ref{vg8a}(b)). 
Since the Wilson loop operator is affected   
by the topological flux through it only mod $N$, this should 
optimize the expectation. But then the picture and expansion in 
terms of groups of isolated independent vortices (\ref{naive}) 
is no longer applicable.

Similarly, (\ref{ineqII}) leads to 
(\ref{quasilin}) because it is obtained by assuming 
the vortex inside each container as completely isolated 
and independent of all the others. The exact expression 
(\ref{W5}) holds for any number of factors in the product 
inside the expectation (reflecting mod $N$ conservation).     
In view of the above discussion, 
one may as well combine containers into ones as thick as 
possible by integrating over the boundary fields of neighboring 
containers. With $T\gg R$, this amounts to taking 
containers in the product in (\ref{W5}), 
(\ref{ineqII}) having   
transverse area $\sim R^2$, and longitudinal extension  
$\sim (\mbox{const}\,R)^{(d-2)}$. 
(\ref{ineqII}) now gives 
\beq
V(R)  \geq \mbox{const}\,R - {\mbox{const}\over R}\,(\,\ln R + 
\mbox{const}\,) \label{strictlin}
\eeq 
replacing (\ref{quasilin}). For loops with $T
\stackrel{\ }{\stackrel{\tx >}{ \sim}}R$, 
basically only one vortex container is needed, which 
means that strict linear potential arises  
essentially from thick vortex fluctuations spanning the 
entire loop area. It is interesting to note that such thick 
vortices also produce nonperturbative $1/R$ contributions  
(at scales outside the short distance perturbation theory regime).   

Inequality (\ref{ineqII}) is actually rather crude. The 
inequalities (\ref{ineqI}), (\ref{chess}) following 
from reflection positivity are much more powerful because 
of the exponents that allow estimates uniform in the lattice size. 
They give directly pure area law. As is easily seen,  
this is true even if one further crudely bounds the r.h.s. 
of (\ref{chess}) from above as done in (\ref{ineqII}), since 
$\ln L_\mu/ A^{\prime} \to 0$ as $|\Lambda|\to \infty$. 
There is no real reason for doing this though.\footnote{ 
(\ref{chess}), however, might serve as a starting point for 
more sophisticated approximations where the behavior (\ref{vfe1}) 
for the vortex free energy under spreading of flux is 
at least partially derived rather than assumed.} 
 
\section{Long range vortex fluctuations -- effective theory} 

The above discussion, based on exact relations and inequalities 
between the Wilson loop and the free energy order parameters, 
indicates that an effective picture of the long distance 
confining fluctuations as isolated, independent vortices 
winding over long distances -  
in other words, as some kind of an approximately dilute or weakly 
interacting vortex gas - is not generally applicable. 
It does not take properly into account the relation between length and 
thickness of a vortex implied by the cost in free energy 
necessary to create the vortex in the first place, nor the 
correlations between vortices caused by their  
long range (topologically nontrivial) pure gauge tails.    
These correlations are present irrespective of the separation and  
enforce the mod N conservation of topological flux. 
Even though the cost diminishes exponentially with the 
transverse thickness of a vortex (creation of mass gap), these 
effects must still be properly accounted for if vortices of basically 
arbitrary length are to be present in the vacuum.  
These effects then generally tend to cause neighboring thick 
vortices to thicken further 
and merge since this lowers the free-energy cost for the vortices 
(above any background of fluctuations that may be present) for 
the same mod N total flux in the system. 
Thus, our previous discussion  
implies that the long distance linear potential should not 
properly be viewed as 
arising from the fluctuations caused by 
a gas of independent vortices winding through the loop, 
each of thickness much less than the linear dimensions of the 
loop. Rather, the fluctuation is 
more accurately described as that due to vortices of 
thickness comparable to the (shorter) linear loop dimension. 
Negative values of the loop 
occur then with almost exactly equal measure weight as positive values 
(no vortex (mod 2)). This optimizes the expectation to be as small as 
possible, i.e. exhibit exact area law.\footnote{Recall that it is a 
rigorous result that the potential cannot rise faster than linearly.}

This reflects the striking behavior revealed by the  
numerical simulations \cite{KT1}, that over 
sufficiently large scales there is `condensation' of   
vorticity carrying flux. Over a hypercube of side length 
of about  1 fm, the weighted probability at large $\beta$ 
for nonzero (mod N) flux configurations goes to unity.  
The weighted probability then that one finds a vortex of at least 
this transverse size going through a 2-dim face on the hypercube 
boundary is approaching one. More generally, above this 
scale, vortices of any length, by corresponding appropriate 
adjustment in thickness, can occur at practically zero free energy cost. 
One may view this as 
percolation of vortices in the following sense. If one 
considers any two disjoint segments on the boundary of 
a large 4-dimensional simply connected region, the probability of 
being joined by a vortex of sufficient thickness ($\geq$ 1 fm) 
is finite. 

This picture of `percolated' vortices in all possible $[\mu\nu]$ 
orientations, with flux in intersections being conserved only 
mod 2, implies that in general it is difficult to unambiguously 
identify individual vortices.\footnote{The way embedded non-self-interacting 
2-dimensional surfaces (in our case, 
surfaces of a certain thickness) can grow to densely fill 4-dimensional 
space, the so called capped gropes and towers, has been 
extensively studied in manifold theory \cite{FQ}.} 
Rather, in the absence of obstructions or boundaries introduced by  
external probes, one can talk about 
an average nonvanishing vorticity field 
defined on the coarse scale of 1 fm, measured by the `circulation' 
(plaquettes, Wilson loops) above this scale.

What simple effective theory can describe this vacuum? We stress 
that we mean an effective theory strictly of only these long range 
vortex fluctuations (confiners) resulting from 
integrating out to an appropriate scale. Even at that scale 
there will of course be all kinds of other $SU(2)$ ($SU(N)$) 
fluctuations which we consider irrelevant for 
confinement. Let us list the minimal requirements 
for the effective theory: 
\begin{enumerate}
\item[(i)] On a coarse scale of about $l=1 - 1.2$ fm, 
the partition function should be expressible solely in 
terms of vortex excitations (coclosed 
(closed dual) surfaces of codim 2). 

\item[(ii)] The mod 2 (mod $N$) property should be manifestly 
incorporated.  
 
\item[(iii)] The vortex flux through each coarse scale 
plaquette should incur an action 
equal to the vortex free energy per unit length for  
thickness $\sim l$, as defined and computed from (\ref{vfe}). 
This amount of free energy for the confiners (their action on the 
coarser scale) is the cost {\it above} the sea of all other 
vacuum fluctuations. (Again, note that 
this depends only mod 2  on the number of `vortex-introducing' 
singular gauge transformations injected in the box in (\ref{vfe}).) 

\item[(iv)] On this coarse lattice, the thick vortices should be in 
a `percolation phase'. Thus, despite (i), 
there should not be   
a useful expansion of the theory -- i.e. a convergent, or,  
at least formally, systematic expansion scheme 
allowing computation of observables -- such that each term in 
the expansion is characterized by a well-defined number of 
vortex excitations. 

\item[(v)] The Wilson loop expectation should give strict 
area law as in the bound (\ref{ineqI}).  
\end{enumerate}

Let then $\Lambda_c$ be the coarse lattice of lattice spacing $\sim l$, 
and $\chi_p \in Z(2)$ variables residing on its plaquettes. 
Then the simplest way to implement (i) above is by the 
partition sum: 
\beq 
Z_{\Lambda_{c}} = \int_{Z(2)}\,\prod_p\,d\chi_p \,\prod_c \,
{\tx 1\over 2}\,\Big(\,1 + \prod_{p\in \partial c} \chi_p\,\Big)  
\;\exp\,A_{\rm eff}\,. \label{z2eff1}
\eeq 
The measure enforces the constraint 
\beq
\prod_{p\in \partial c} \chi_p =1 \label{z2con}
\eeq
on the plaquettes forming the boundary of every 3-dimensional 
cube $c$ on $\Lambda_c$, so only 
excitations on coclosed sets are allowed. Equivalently, 
on the dual lattice, (\ref{z2con}) assumes the form 
\beq
\prod_{p\in \partial^* b} \chi^*_p =1 \label{z2con*}
\eeq
on the plaquettes forming the coboundary of every bond. 
The requirement (ii) is then automatically taken care of.

The general form of the effective action $A_{\rm eff}$   
\beq
A_{\rm eff} = \beta_{\rm eff}\,\sum_p \chi_p\, +\, \beta_{2p}\!\!
\sum_{(p,p^{\prime})\in \partial^*b} \chi_p\,\chi_{p^\prime} 
\,+\, \beta_{3p}\!\!
\sum_{(p,p^{\prime},p^{\prime\prime})\in \partial^*b} 
\chi_p\,\chi_{p^\prime}\,\chi_{p^{\prime\prime}} + \cdots 
\label{effact}
\eeq
involves, in addition to the basic plaquette term, quasilocal 
interaction terms involving two or more plaquettes in the 
coboundary of each bond, etc. 
Now, from (iii), and (\ref{vfe1}), we must have 
\beq
\beta_{\rm eff} \sim \exp(\,-\rho(\beta)\,l^2\,) \label{betaeff}
\eeq  
giving, in principle, the coupling $\beta_{\rm eff}$ 
in terms of the coupling at the original lattice spacing, as 
$\rho$ must approach the string tension for sufficiently 
large $l$. From the numerical simulations \cite{KT1}, $l \sim$ 1.1 fm.  
This gives $\beta_{\rm eff} \sim 0.002$. This very small value 
reflects of course the fact that at this  choice of the physical 
length $l$  vortex flux is found to become very `light'. 
Correspondingly, the terms involving 
products of two or more plaquettes must be of 
order $\beta_{\rm eff}^2$ and higher, 
hence entirely negligible.  

The effective model ({\ref{z2eff1}) is now seen to simply be 
a $Z(2)$ LGT. Indeed, the constraint (\ref{z2con}) can be 
explicitly solved by introducing $Z(2)$ bond variables 
$\gamma_b$ by: 
\beq
\chi_p = \prod_{b\in \partial p}\;\gamma_b \,.\label{z2solv}
\eeq
Then 
\beq
Z_{\Lambda_{c}} = \int_{Z(2)}\,\prod_b\,d\gamma_b   
\;\exp\Big(\beta_{\rm eff}\,\sum_p \gamma_{\partial p} + 
\cdots \,\Big)\,, 
\label{z2eff2}
\eeq
where the ellipses indicate the additional clover and higher loop  
terms corresponding to the additional terms in (\ref{effact}). 
The theory is in the deep strong coupling regime 
$\beta_{\rm eff} \ll 1$. Thus (iv) above is indeed 
satisfied. The theory can be treated in the strong coupling expansion. 
It cannot, however, be meaningfully expanded in its vortex 
excitations -- that would be appropriate in the weak coupling 
$\beta _{\rm eff} \gg 1$  regime in the form of the usual 
weak coupling expansion for discrete groups (dilute vortex gas) 
\cite{MMS}.    

It should perhaps be explicitly pointed out that the $Z(2)$ 
variables in (\ref{z2eff1}), (\ref{z2eff2}), serve as an effective  
description of long distance fluctuations creating 
topological $Z(2)$ flux 
(elements of $\pi_1(SU(2)/Z(2)$) in the original theory \cite {KT2}.  
They have nothing to do with the $Z(2)$ part of the original 
$SU(2)$ bond variables. Note that the  
$Z(2)$ gauge theory interaction, together with (\ref{betaeff}), 
correctly reproduce the effects of 
flux spreading and thickening of vortices while conserving 
flux mod 2. Correspondingly, the Wilson loop now 
automatically gives the correct area law.

The Wilson loop in (\ref{z2eff2}) 
represents the coupling of an external quark 
current to the long distance confining fluctuations. 
Its replacement by dynamical quarks introduces  
fermions in the medium of these fluctuations.  
Since the effective coupling is strong, it induces   
dynamical chiral symmetry breaking (CSB).    

CSB in strongly coupled LGT has actually been demonstrated  
analytically in the superstrong gauge coupling 
limit (no plaquette action term) by expansion or mean field 
(large $N$ or $d$) approximations \cite{B}, \cite{KS-Ketal}, 
and rigorously by infrared bounds \cite{SS}. It is 
physically obvious that the result extends to a finite region 
in the strong coupling regime.\footnote{It should be possible to 
prove this by cluster expansion techniques around the $\beta=0$ 
point, though not so straightforward for technical reasons  
(bounding terms with Grassmann integrands).} 

We may try to use the results in \cite{B} - \cite{SS} to get 
an estimate of the contribution to the quark condensate 
in the effective theory. Corrections from the plaquette 
term in the action are totally 
negligible due to the smallness of $\beta_{\rm eff}$. 
One then has: 
\beq
\vev{\bar{q}q} = z(l) N\,{1\over l^3}\,\sqrt{{2\over d}}\Big(\,
1-k(d)\,\Big)^{1/2}\,, \label{chconds}
\eeq
where ${1\over 8}\leq k(4) < 0.35$. $z(l)$ is some renormalization 
factor that, in a more sophisticated treatment, should depend 
on how fermions are introduced at the original lattice spacing. 
Here we naively set it equal to one -- this is equivalent to 
simply taking staggered fermions on the coarse lattice.  
With $l = $ 1.1 fm, this gives $\vev{\bar{q}q}= (195 \ {\rm MeV})^3$  
for $N=2$, and $(223 \ {\rm MeV})^3$ for $N=3$. This indicates  
that the quark condensate may be entirely accounted 
for by the long range confining fluctuations.

Following our previous development, the effective $Z(N)$ 
theory appears to emerge rather naturally, and in fact in a fairly unique 
manner. The idea that an effective theory of long range vortex fluctuations 
must be a $Z(N)$ LGT is not new, but has not, we believe, been 
formulated in this way before.   
Recently, a model equivalent to (\ref{z2eff1}), 
in the representation (\ref{z2con*}) 
and employing additive $Z(2)$ variables, was considered in 
\cite{Tu}, apparently without any reference to  
$Z(2)$ LGT.

\section{Conclusions}

In the present paper we studied the energetics of how 
vortices can disorder Wilson loops of different sizes.
Vortices of any thickness smaller than the linear size
of a given Wilson loop can link with it and contribute to
disordering its average. Here we pointed out that for
Wilson loops of any given size it is the vortices of 
``maximal'' thickness, i.e.\ thickness of the order of
the linear size of the loop, that give the
most important contribution resulting in an area-law 
suppression of large Wilson loops and a linear heavy quark
potential. On the other hand, vortices of any {\em fixed} thickness 
contribute only with a logarithmically suppressed term to the potential.
This is in contradiction with the naive picture of 
confinement by vortices which assumes that vortices of
fixed thickness can link with a fixed probability 
with arbitrarily large Wilson loops. 
The correct picture must take properly into account the 
relation between the length and thickness of a vortex imposed by the 
free energy requirements for creating the vortex, as well as the 
interaction between vortices introduced by the constraint 
of mod N conservation of the vortex flux.  
This picture of vortices naturally yields a long distance 
effective $Z(N)$ gauge theory above the confinement scale
of around 1fm. The effective theory is deep in the 
strong coupling regime which makes it impossible to 
interpret it in terms of a simple vortex gas expansion. The only
useful expansion one can consider is the strong coupling one.
Being deep in the strong coupling regime, the effective
$Z(N)$ gauge theory naturally produces chiral symmetry breaking 
in the presence of fermion fields.

\end{document}